\documentclass[aps, prb, reprint, superscriptaddress]{revtex4-1}
\usepackage{amssymb}
\usepackage{amsmath}
\usepackage{graphicx}
\usepackage{bm}
\usepackage{color}

\begin{document}

\title{Non-uniform heat redistribution among multiple channels\\
in the integer quantum Hall regime}
\author{Ryota Konuma}
\author{Chaojing Lin}
\author{Tokuro Hata}
\author{Taichi Hirasawa}
\affiliation{Department of Physics, Tokyo Institute of Technology, 2-12-1 Ookayama,
Meguro, Tokyo, 152-8551, Japan.}
\author{Takafumi Akiho}
\author{Koji Muraki}
\affiliation{NTT Basic Research Laboratories, NTT Corporation, 3-1 Morinosato-Wakamiya,
Atsugi 243-0198, Japan.}
\author{Toshimasa Fujisawa}
\email{fujisawa@phys.titech.ac.jp}
\affiliation{Department of Physics, Tokyo Institute of Technology, 2-12-1 Ookayama,
Meguro, Tokyo, 152-8551, Japan.}

\begin{abstract}
Heat transport in multiple quantum-Hall edge channels at Landau-level
filling factor $\nu $ = 2, 4, and 8 is investigated with a quantum point
contact as a heat generator and a quantum dot as a local thermometer. Heat
distribution among the channels remains highly non-uniform, which can be
understood with the plasmon eigenmodes of the multiple channels. The heat
transport can be controlled with another quantum point contact as a
quantized heat valve, as manifested by stepwise increases of heat current at
the thermometer. This encourages developing integrated heat circuits with
quantum-Hall edge channels.
\end{abstract}

\date{\today }
\maketitle

\section{Introduction}

Recent developments in quantum thermoelectric devices, such as heat engines
and refrigerators based on quantum-dot (QD) and superconducting circuits,
are attractive for manipulating heat in beneficial ways. \cite%
{BookGemmer,ThierschmannNatNano2015,KoskiPRL2015,TanNatComm2017,LindenfelsPRl2019,JalielPRL2019,KOnoPRL2020,DuttaPRL2020,JosefssonNatNano2018,HarzheimPRR2020}
Even though each device works efficiently, the ordinary heat diffusion
process between these devices can degrade the overall thermal efficiency
dramatically.\cite{JosefssonNatNano2018,HarzheimPRR2020} This issue can be
overcome by using chiral heat transport in quantum Hall (QH) edge channels.
In the integer QH regime, heat travels unidirectionally in a certain
direction determined by the direction of the magnetic field.\cite%
{SanchezPRL2015,SamuelssonPRL2017,Sanchez-PRL2019,HajilooPRB2020,Borin-PRB2019}
Each channel has a quantized thermal conductance as well as a quantized
electronic conductance.\cite{Jezouin-Science2013} For typical edge channels
in AlGaAs/GaAs heterostructures, the heat is mostly conserved in the
channels within a dissipation length of about 30 $\mu $m.\cite%
{leSueurPRL2010,Itoh-PRL2018} These characteristics are attractive for
designing an integrated heat circuit, where functional thermoelectric
devices are efficiently connected by QH edge channels. Fundamental heat
transport in nanostructures, such as thermal Coulomb blockade, has been
successfully studied.\cite{SivreNatPhys2017,Sivre-NatCom2019,DuprezPRR2021}

Besides the fundamental heat transport characteristics, QH edge channels
have unique characteristics of a Tomonaga-Luttinger (TL) liquid as
extensively studied at Landau-level filling factor $\nu $ = 2.\cite%
{BookGiamarchi,LevkivskyiBook2012,Hashisaka-RevPhys2018,FujisawaAnnPhys2022}
When one of the two channels is heated with a quantum point contact (QPC),
the Coulomb interaction redistributes the charge and heat in the two
channels.\cite%
{BergPRL2009,FreulonNatComm-SC,Bocquillon-NatCom2013,Inoue-PRL2014,Hashisaka-NatPhys2017}
This heat redistribution occurs in a deterministic way in terms of
spin-charge separation. Because this process can be understood as plasmon
scattering at discontinuities, i.e., the charge injection points for the
above cases, the electronic state may remain in a non-thermal metastable
state during the transport in clean channels.\cite%
{LevkivskyiPRB2012,Itoh-PRL2018,WashioPRB2016} This is not the case for a
diffusion process with stochastic scattering events.\cite{PothierThermalize}
One can utilize such QH channels for transferring heat from one functional
device to another efficiently without maximizing the entropy of the
carriers. While the two-channel configuration at $\nu $ = 2 is the
archetypal case for the TL model, one can also expect similar behavior for
multiple channels at higher $\nu $. Observation of non-uniform
heat redistribution among the channels would justify the applicability of
the TL physics.

In this work, we investigate heat transport through multiple QH edge
channels from a QPC heat generator to a QD thermometer at $\nu $ = 2, 4, and
8 in an AlGaAs/GaAs heterostructure. When a particular channel is heated
with the QPC, the heat is redistributed among the copropagating channels in
a non-uniform way independent of the propagation distance. This is
consistent with the plasmon modes of the TL liquid with multiple channels.
We also demonstrate that the heat transport can be controlled with another
QPC as a heat valve placed in the edge channels, as manifested by stepwise
increases of heat current at the QD thermometer. Our experiment encourages
developing integrated heat circuits with QH edge channels in integrated heat
circuits.

\begin{figure}[tbp]
\begin{center}
\includegraphics[width = 3.3in]{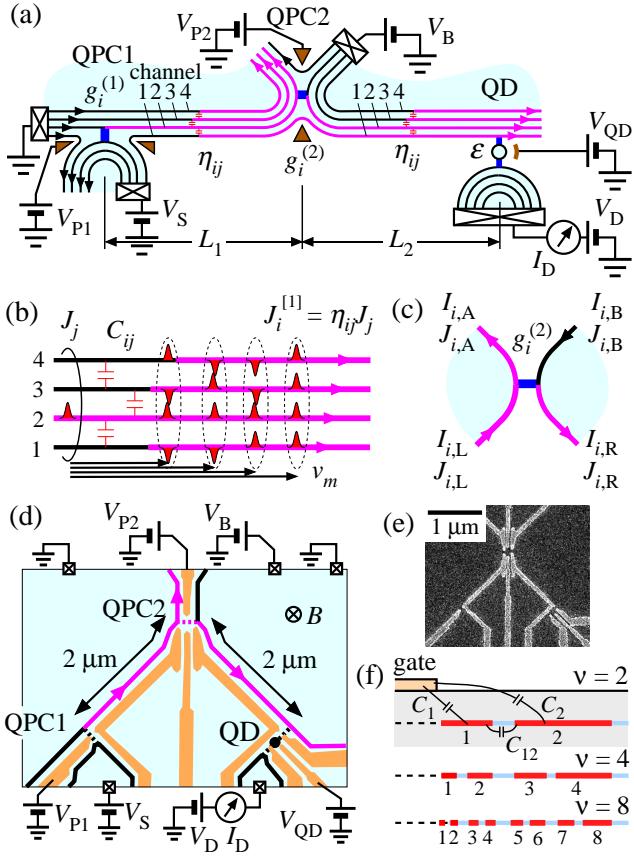}
\end{center}
\caption{(a) Schematic of the heat transport measurement with QPC1 as a heat
generator, QPC2 as a heat valve, and QD as a heat detector. Four QH edge
channels 1 - 4 with chirality depicted by the arrows are heated in the
magenta regions. (b) Fractionalization process in the 4 channels when a
charge pulse is introduced in channel 2. Plasmon modes with coupled
wavepackets in ovals carry heat. (c) Partition process at a QPC for heat
current $J_{i,p}$ and charge current $I_{i,p}$ in channel $i$ at port $p\in
\left\{ \mathrm{L,R,A,B}\right\} $. (d) Schematic device structure with a
measurement setup. (e) Scanning electron micrograph of a control sample. (f)
Schematic channel geometries at $\protect\nu $ = 2, 4, and 8.}
\end{figure}

\section{Heat circuit}

Figure 1(a) shows a schematic diagram of edge channels for studying heat
transport at $\nu $ = 4. Four chiral channels with index $i$ (= 1, 2, 3, and
4 from the outer to the inner one) are formed along the edges of the
incompressible regions (light blue) under perpendicular magnetic field $B$.
Heat can be introduced from the first QPC (QPC1) to the left segment of the
target channels of length $L_{1}$ . This heat is redistributed in the four
channels during the transport and partitioned at the second QPC (QPC2). The
transmitted heat is redistributed again in the right segment of length $%
L_{2} $, and the resulting electronic excitation is investigated by a QD
spectrometer. With this setup, we demonstrate tunable heat transport through
QPC2 as a heat valve. The operation principle of each device is described in
the following before showing the experimental results.

By applying a source voltage $V_{\mathrm{S}}$ across QPC1, charge and heat
currents can be introduced to channel $i$ with transmission coefficient $%
g_{i}^{(1)}$. By assuming energy independent $g_{i}^{(1)}$ for simplicity,
electrochemical potential $\mu _{i}=eg_{i}^{(1)}V_{\mathrm{S}}$, charge
current $I_{i}=\frac{e^{2}}{h}g_{i}^{(1)}V_{\mathrm{S}}$, and heat current $%
J_{i}=\frac{1}{2h}g_{i}^{(1)}\left( 1-g_{i}^{(1)}\right) V_{\mathrm{S}}^{2}$
are induced in the channels, where $e$ is the elementary charge and $h$ is
the Planck constant.\cite{BookHeikkila,Sivre-NatCom2019} For example, Fig.
1(a) shows that the heat is introduced from QPC1 to channel 2 with $%
g_{1}^{(1)}=1$, $g_{2}^{(1)}=0.5$, and $g_{3}^{(1)}=g_{4}^{(1)}=0$.

In the presence of inter-channel Coulomb interaction, the heat current is
redistributed during the transport in the copropagating channels. The
inter-channel tunneling is weak in the following experiment and thus
neglected for simplicity. The heat current in one-dimensional channels can
be expressed as an ensemble of plasmons (charge density waves), which
describe electronic excitation from the ground state.\cite{WashioPRB2016} In
the low-energy limit, the plasmons can be described as superpositions of the
transport eigenmodes determined by the Coulomb interaction. This description
was successful in studying heat transport and non-thermal states at $\nu $ =
2.\cite{Safi-PRB1995,Hashisaka-PRB2012,HashisakaPRB2013,FujisawaPRB2021} We
apply this scheme to higher $\nu $ (2 - 8) in this work. Figure 1(b)
illustrates how a single charge wave packet generated in $i$ = 2 channel
splits into four coupled wavepackets (enclosed by dashed ovals) propagating
at different velocities $v_{m}$ for eigenmode $m$. This is a kind of
fractionalization in which an electron wave packet from a point contact is
fractionalized into several quasiparticles. By considering the initial
packet of time width $\Delta t=\hbar /eV_{\mathrm{S}}$, the
fractionalization should develop after traveling the fractionalization
length $L_{\mathrm{F}}=\max\limits_{m\neq n}\left\{ \frac{v_{m}v_{n}}{%
\left\vert v_{m}-v_{n}\right\vert }\Delta t\right\} $, as derived in
Appendix. Such electronic wave packets randomly injected from a QPC are
redistributed in all channels, which is the microscopic picture of the heat
redistribution. By introducing the heat fractionalization factor $\eta _{ij}$
from channel $j$ to channel $i$, the heat current in channel $i$ after the
fractionalization can be written as $J_{i}^{\left[ 1\right] }=\sum_{j}\eta
_{ij}J_{j}$, where the superscript [1] denotes the first fractionalization
process. The plasmon scattering model suggests reciprocal
relation $\eta _{ij}=\eta _{ji}$, as shown in Appendix. This
factor is determined by the strengths of inter-channel Coulomb interactions
and should be different from equipartition ($\eta _{ij}=1/N$ for $N$
channels) expected in ordinary thermal conduction associated with the
diffusive motion of electrons or phonons in materials.

The heat transport can be controlled by using QPC2 with transmission
coefficient $g_{i}^{(2)}$ for channel $i$. Considering that the electronic
states in the channels are correlated as depicted in Fig. 1(b), the TL
characteristics such as the power-law anomaly in the conductance may appear
in this tunneling process. Here, we shall ignore it for simplicity. This may
be justified if $g_{i}^{(2)}$ is not very close to 0 and 1, where previous
measurements with QD spectroscopy did not find the non-thermal states
associated with the TL liquid.\cite{Itoh-PRL2018} Therefore, we assume that
each electron in the input channels is allowed to tunnel stochastically to
the output channels. This partition process for $i$-th channel is
schematically shown in Fig. 1(c) for two input (L and B) and two output (A
and R) ports with finite $g_{i}^{(2)}$. One can show how heat currents $J_{i,%
\mathrm{L}}$ and $J_{i,\mathrm{B}}$ as well as charge currents $I_{i,\mathrm{%
L}}=\frac{e}{h}\mu _{i,\mathrm{L}}$ and $I_{i,\mathrm{B}}=\frac{e}{h}\mu _{i,%
\mathrm{B}}$ with chemical potentials $\mu _{i,\mathrm{L}}$ and $\mu _{i,%
\mathrm{B}}$ will be partitioned with $g_{i}^{(2)}$ for arbitrary
single-particle distribution functions. The output charge current $I_{i,%
\mathrm{R}}$ and heat current $J_{i,\mathrm{R}}$ are given by%
\begin{equation}
I_{i,\mathrm{R}}=g_{i}^{(2)}I_{i,\mathrm{L}}+\left( 1-g_{i}^{(2)}\right)
I_{i,\mathrm{B}}  \label{IiR}
\end{equation}%
\begin{eqnarray}
J_{i,\mathrm{R}} &=&g_{i}^{(2)}J_{i,\mathrm{L}}+\left( 1-g_{i}^{(2)}\right)
J_{i,\mathrm{B}}  \label{JiR} \\
&&+\frac{1}{2h}g_{i}^{(2)}\left( 1-g_{i}^{(2)}\right) \left( \mu _{i,\mathrm{%
L}}-\mu _{i,\mathrm{B}}\right) ^{2}  \notag
\end{eqnarray}%
where energy independent $g_{i}^{(2)}$ is assumed for simplicity. Equation (%
\ref{JiR}) shows that the heat partition at QPC2, described in the first and
second terms, can be controlled with $g_{i}^{(2)}$, while the heat
generation at the QPC2, given by the third term, can be zeroed by setting $%
\mu _{i,\mathrm{L}}=\mu _{i,\mathrm{B}}$. The latter is particularly useful
to discriminate the heat partition from heat generation at QPC2.

The heat partitioned at QPC2 is redistributed in the right segment of the
target channels with the same $\eta _{ij}$ as the left one. The resulting
heat current in channel $i$ after the second partition should be given by $%
J_{i}^{\left[ 2\right] }=\sum_{j}\eta _{ij}J_{j,\mathrm{R}}$ from the above
assumptions. We can experimentally investigate $J_{1}^{\left[ 2\right] }$ in
the outermost channel by using a QD energy spectrometer.\cite%
{Altimiras-NatPhys10} For a single energy level $\varepsilon $ in the QD
weakly coupled to the channel, the current $I_{\mathrm{D}}=e\Gamma \left[
f_{1}\left( \varepsilon \right) -f_{\mathrm{D}}\left( \varepsilon \right) %
\right] $ with the single-electron tunneling rate $\Gamma $ measures the
tunneling distribution function $f_{1}\left( \varepsilon \right) $ %
of interest if the distribution function $f_{\mathrm{%
D}}\left( \varepsilon \right) $ of the drain electrode is known. The
chemical potential $\mu _{1}=\int_{-\infty }^{\infty }\left[ f_{1}\left(
E\right) -f_{\mathrm{FD}}\left( E;\mu =0,T=0\right) \right] dE$ and heat
current $J_{1}=\frac{1}{h}\int_{-\infty }^{\infty }E\left[ f_{1}\left(
E\right) -f_{\mathrm{FD}}\left( E;\mu _{1},T=0\right) \right] dE$ can be
obtained from $f_{1}\left( \varepsilon \right) $, where $f_{\mathrm{FD}%
}\left( E;\mu ,T\right) $ is the Fermi distribution function at chemical
potential $\mu $ and temperature $T$. If $f_{1}\left( \varepsilon \right) $
is approximated with $f_{\mathrm{FD}}\left( E;\mu _{1},T_{1}\right) $, the
corresponding heat current $J_{1}=\frac{\pi ^{2}}{6h}k_{B}^{2}T_{1}^{2}$ can
be estimated from the effective temperature $T_{1}$ of the channel.

While energy-independent $g_{i}$ is assumed in the above analysis, the
transmission coefficient can be energy-dependent in an actual device. In
this case, $g_{i}$ can be understood as the average coefficient for the
energy range of interest for $\mu _{i}$ and $I_{i}$. Importantly, the
energy-dependent tunneling induces thermoelectric effects, which modify Eqs.
(\ref{IiR}) and (\ref{JiR}). For example, different input temperatures with $%
J_{i,\mathrm{L}}\neq J_{i,\mathrm{B}}$ causes thermoelectric potential ($\mu
_{i,\mathrm{R}}\neq \mu _{i,\mathrm{A}}$) in the outputs even when no
potential bias is given in the inputs ($\mu _{i,\mathrm{L}}=\mu _{i,\mathrm{B%
}}$). Even in the absence of input heat currents ($J_{i,\mathrm{L}}=J_{i,%
\mathrm{B}}=0$), the heat generation of the third term of Eq. (\ref{JiR}) is
modified to cause asymmetric heat currents in the outputs ($J_{i,\mathrm{R}%
}\neq J_{i,\mathrm{A}}$). Such non-linear effects will be discussed later
with the experiment.

\section{Experimental results}

\subsection{Device}

The above heat transport was investigated with a device fabricated in a
standard AlGaAs/GaAs heterostructure with an electron density of 1.85$\times 
$10$^{11}$ cm$^{-2}$ and electron mobility on the order of 10$^{6}$ cm$^{2}$%
/Vs at low temperature. A schematic device structure and a scanning electron
micrograph of a control device are shown in Figs. 1(d) and 1(e),
respectively. Under perpendicular magnetic field, $B$ = 3.3 T ($\nu \simeq $
2), 1.8 T ($\nu \simeq $ 4), and 0.9 T ($\nu \simeq $ 8), QPC1, QPC2, and QD
were formed by applying appropriate voltages on the gates. We
restricted ourselves to these even filling factors to ensure\ a large energy
gap in the bulk. All measurements were performed at a base
temperature of about 30 mK, and the QD thermometry shows an effective
electron temperature of about 170 mK in the absence of QPC heating %
(due to external noise in the experimental environment).

\begin{figure}[tbp]
\begin{center}
\includegraphics[width = 3.3in]{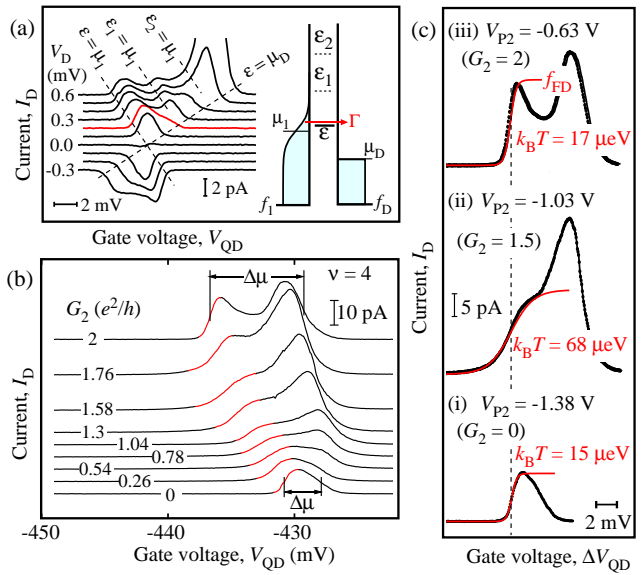}
\end{center}
\caption{ (a) Single-electron transport characteristics of the QD at $B$ =
1.8 T ($\protect\nu =$ 4). The inset shows the energy diagram with the dot
levels, $\protect\varepsilon $, $\protect\varepsilon _{1}$, and $\protect%
\varepsilon _{2}$. The QD spectroscopy was performed with small bias
typically at $V_{\mathrm{D}}=$ 0.2 mV (the red trace). (b) QD current
spectrum at $V_{\mathrm{D}}=$ 0.2 mV obtained when the current and heat were
introduced from QPC2 at $V_{\mathrm{S}}=$ -0.3 mV with several conductance $%
G_{2}$ values. The current onset in red is broadened at $G_{2}\sim $ 0.5 and
1.5. The width of the conductive region measures $\Delta \protect\mu $. (c)
The current profiles (black lines) fitted with the Fermi distribution
function (red lines) around the onset.}
\end{figure}

Figure 2(a) shows the Coulomb blockade characteristics of the QD obtained at 
$B$ = 1.8 T ($\nu \simeq $ 4), where both QPCs were fully open for reference
and the QD was biased by varying $V_{\mathrm{D}}$ at $V_{\mathrm{S}}$ = 0.
The current $I_{\mathrm{D}}$ increases stepwise when energy levels $%
\varepsilon $ of the ground state and $\varepsilon _{1}$ and $\varepsilon
_{2}$\ of the first and second excited states enter the transport window
between the chemical potentials $\mu _{1}$ (= 0) and $\mu _{\mathrm{D}}=eV_{%
\mathrm{D}}$. The stepwise feature is pronounced with a peak associated with
the Fermi-edge singularity.\cite{FrahmPRB2019} This many-body effect is
smeared out when the heat is introduced to the channel and thus neglected in
the following analysis. For all the spectroscopy measurements presented
below, we apply small but finite energy bias $\Delta \mu =\mu _{1}-\mu _{%
\mathrm{D}}=$ 0.1 -- 0.5 meV to allow us to investigate the distribution
function $f_{1}\left( E\right) $ at around $\varepsilon \sim \mu _{1}$.

\subsection{Heat redistribution}

First, we investigate heat redistribution among copropagating edge channels,
which is required for analyzing the heat-valve characteristics. To confirm
that the heat redistribution is independent of the propagating length, we
compare two experiments with short ($L_{2}=$ 2 $\mu $m) and long ($%
L_{1}+L_{2}=$ 4 $\mu $m) distances from QPC2 and QPC1, respectively, to the
QD. For the short configuration, the charge and heat were introduced from
QPC2 at various conductance $G_{2}=\sum_{i}g_{i}^{(2)}$, with $V_{\mathrm{S}%
}=$ -0.3 mV and QPC1 fully opened. QD current profiles were obtained with $%
V_{\mathrm{D}}=$ 0.2 mV, as shown in Fig. 2(b). The width of the conductive
region increases with $G_{2}$, from which $\Delta \mu =\mu _{1}-\mu _{%
\mathrm{D}}$ can be estimated. The onset of the current step at around $%
\varepsilon \sim \mu _{1}$, which is highlighted in red, is broadened when
the heat is injected to channel 1 at $G_{2}\sim $ 0.5 and channel 2 at $%
G_{2}\sim $ 1.5. The latter demonstrates the heat transfer from channel 2 to
channel 1, meaning that the heat is redistributed between channels 1 and 2.
The current profile around the onset can be fitted with the Fermi
distribution function, as shown in Fig. 2(c). The effective temperature of $%
k_{\mathrm{B}}T=$ 68 $\mu $eV estimated for $G_{2}\sim $ 1.5 is
significantly greater than $k_{\mathrm{B}}T=$ 15 - 17 $\mu $eV for $G_{2}$ =
2 and 0 that corresponds to the base temperature. Corresponding heat current 
$J=\frac{\pi ^{2}}{6h}k_{B}^{2}T^{2}$ ($\simeq $ 0.29 pW at $G_{2}\sim $
1.5) in channel 1 is obtained from the quantized heat conductance.

\begin{figure}[tbp]
\begin{center}
\includegraphics[width = 3.3in]{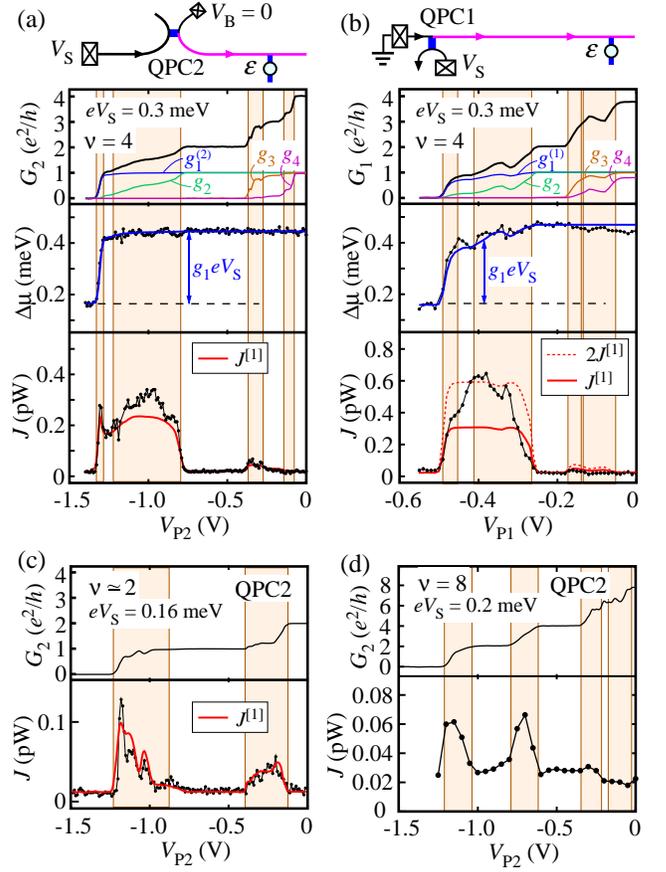}
\end{center}
\caption{Summary of single QPC experiments showing conductance $G_{1}$ or $%
G_{2}$ in the top panels, $\Delta \protect\mu =\protect\mu _{1}-\protect\mu %
_{\mathrm{D}}$ in the middle panels, and the heat current $J$ in the bottom
panels. (a) excited from QPC2 at distance $L_{2}=$ 2 $\protect\mu $m and $%
\protect\nu $ = 4, (b) excited from QPC1 at distance $L_{1}+L_{2}=$ 4 $%
\protect\mu $m and $\protect\nu $ = 4, (c) excited from QPC2 at %
$\protect\nu \simeq $ 2, and (d) excited from QPC2 at $\protect%
\nu $ = 8. The tunneling regions at non-integer $G$ are highlighted with
orange stripes. Schematic channel geometries with QPC1 and QPC2 are shown in
the insets to (a) and (b).}
\end{figure}

Figure 3(a) summarizes the gate voltage $V_{\mathrm{P2}}$ dependence of the
QPC2 conductance $G_{2}$ in the top panel, $\Delta \mu =\mu _{1}-\mu _{%
\mathrm{D}}$ estimated from the width of the current peak in the middle
panel, and the heat current $J$ in the bottom panel. The tunneling regions
with non-integer $G$ are highlighted by four orange stripes. $\Delta \mu $
increases stepwise only in the leftmost stripe for the tunneling into
channel 1. $\mu _{1}$ increases to $eV_{\mathrm{S}}$ as soon as $%
G_{2}$ reaches 1, and stays almost constant ($\mu _{1}\simeq eV_{\mathrm{S}}$%
) even when the chemical potential of the second channel, $\mu _{2}$, is
changed in the range of $0<\mu _{2}<eV_{\mathrm{S}}$ ($1<G_{2}<2$). This
indicates that inter-channel tunneling is negligible. However, the heat
behaves differently from the charge. The heat current $J$ in channel 1 is
enhanced in multiple tunneling regions, and the maximum $J$ values are
different for different tunneling regions. As the heat is originally
generated in one of the four channels at the QPC, the heat is redistributed
non-uniformly among the channels during the transport before reaching the QD
spectrometer.  By assuming that identical heat $J_{j}=\frac{1}{%
8h}\left( eV_{\mathrm{S}}\right) ^{2}$ ($=$ 0.43 pW for $eV_{\mathrm{S}}=$
0.3 meV) is generated in the $j$-th channel at $g_{j}^{(2)}=0.5$ in the $j$%
-th stripe, the maximum heat current $J=\eta _{1j}J_{j}$ measures the heat
partition factor $\eta _{1j}$; $\eta _{11}\simeq \eta _{12}\simeq 0.5$, $%
\eta _{13}\simeq 0.06$, and $\eta _{14}\simeq 0$ for the data in Fig. 3(a)
with propagation length $L_{2}$ = 2 $\mu $m. This indicates efficient heat
transfer between channels 1 and 2, while heat transfer from channels 3 and 4
is inefficient. Since the sum is close to unity ($%
\sum\nolimits_{i}\eta _{i1}=$ $\sum\nolimits_{j}\eta
_{1j}\simeq 1$), the heat is well conserved within the multiple channels.

The same experiment was performed for the long distance by varying the
conductance $G_{1}$ of QPC1, while QPC2 was fully opened ($G_{2}=4$), as
shown in Fig. 3(b). The conductance plateaus in $G_{1}$ are less clear as
compared to those for QPC2, and similarly, the stepwise increase of $\Delta
\mu $ in the middle panel is not as sharp as that for QPC2. As a result, the
heat current $J$ in the bottom panel shows a broad peak spread over the
leftmost and second-left tunneling regions. It should be noted that the heat
currents in the third and fourth tunneling regions remain low, comparable to
those for QPC2. This implies that a fixed fraction of the heat generated in
the respective channels comes to channel 1 in agreement with the plasmon
transport model. If the heat redistribution was mediated by a diffusion
process, the heat should be redistributed more uniformly for longer
propagation lengths. Moreover, similar heat redistributions ($\eta _{1j}$)
at 2 and 4 $\mu $m suggest that the fractionalization length $L_{\mathrm{F}}$
is shorter than 2 $\mu $m in agreement with the previous study ($L_{\mathrm{F%
}}\simeq $ 0.3 $\mu $m at $V_{\mathrm{S}}=$ 0.3 mV for $\nu $ = 2).\cite%
{Itoh-PRL2018}

We next compare the fractionalization at different filling factors, %
$\nu \simeq $ 2 ($B$ = 3.3 T) \lbrack Fig. 3(c)]
and $\nu $ = 8 [Fig. 3(d)], with that at $\nu $ = 4 [Fig. 3(a)], for
injection from QPC2. Strongly non-uniform heat distribution is seen in all
cases; $\eta _{11}\simeq 0.7$ and $\eta _{12}\simeq 0.3$ at $\nu =2$ and $%
\eta _{11}+\eta _{12}\simeq \eta _{13}+\eta _{14}\simeq 0.5$, $\eta
_{15}+\eta _{16}\simeq 0.05$, and $\eta _{17}+\eta _{18}\simeq 0$\ at $\nu
=8 $. This might be reasonable by considering that the locations of edge
channels at different $\nu $ are related to each other, as schematically
shown in Fig. 1(f). Namely, if large non-uniformity with $\eta _{11}>\eta
_{12}$ is seen at $\nu =2$, similar or even larger non-uniformity with $\eta
_{11},\eta _{12}>\eta _{13},\eta _{14}$ should appear at $\nu =4$ by
considering that the channels 2 and 3 are greatly separated by the cyclotron
energy. Such non-uniformity can be related to the asymmetric channel
structure with different channel capacitances $C_{1}\neq C_{2}$ for $\nu =2$%
. Similar asymmetry was seen in previous charge-fractionalization
experiments; for example $\eta _{11}\simeq $ 0.55 and $\eta _{12}\simeq $
0.45 at $\nu $ = 2 (see Appendix).\cite{Hashisaka-NatPhys2017}

To justify the above analysis, the expected heat current $J_{i}^{\left[ 1%
\right] }$ is calculated in the following way. Because the plateaus at $G=$
1 and 3 are not well quantized, insufficient spin splitting is considered to
decompose $G_{1}$ and $G_{2}$ to $g_{i}^{(1)}$ and $g_{i}^{(2)}$,
respectively, as shown by the colored lines in Fig. 3(a) and (b). The
inter-channel tunneling is neglected for simplicity. To this end, we use a
saddle point potential to characterize a QPC, which allows us to relate $%
g_{1}$ and $g_{2}$ through a single parameter $\alpha $ as $%
g_{2}^{-1}-1=\alpha \left( g_{1}^{-1}-1\right) $. Here, $\alpha =2\pi \Delta
/\hbar \omega _{x}$ is determined by the energy gap (spin splitting) $\Delta 
$ and the characteristic energy $\hbar \omega _{x}$ of the barrier potential.%
\cite{FertigPRB1987,BookIhn} The parameter $\alpha $ is determined from the
deviation of $\Delta \mu $ from $eV_{\mathrm{S}}$ at $G=1$ ($\alpha \simeq $
16 for QPC1 showing $g_{1}^{(1)}\simeq $ 0.8 and $g_{2}^{(1)}\simeq $ 0.2 at 
$G_{1}=1$, and $\alpha \simeq $ 100 for QPC2 showing $g_{1}^{(2)}\simeq $
0.9 and $g_{2}^{(2)}\simeq $ 0.1 at $G_{2}=1$). This $\alpha $ was used to
decompose $G=g_{1}+g_{2}$ into $g_{1}$ and $g_{2}$ for the data at $G\leq 2$%
, by assuming constant $\alpha $. As the $G=2$ plateau is clear, $g_{3}$ and 
$g_{4}$ should be zero at $G\leq 2$, and $g_{1}$ and $g_{2}$ should be 1 at $%
G\geq 2$. For the data at $2\leq G\leq 4$, the same $\alpha $ was used to
decompose $G=2+g_{3}+g_{4}$ into $g_{3}$ and $g_{4}$, as shown in the
figures. The obtained $g_{1}$ explains the variation of $\Delta \mu $ as
shown by the blue line $g_{1}eV_{\mathrm{S}}$ in the middle panels of Figs.
3(a) and (b). The heat current $J^{\left[ 1\right] }$ is calculated from the
above $g_{i}$ and $\eta _{11}=\eta _{12}=0.47$, $\eta _{13}=0.06$, and $\eta
_{14}=0$, as shown by the red line in the bottom panels in Figs. 3(a) and
3(b). A similar analysis was made for the data in Fig. 3(c),
where the decomposition of $G=g_{1}+g_{2}$ is well defined with a large spin
splitting at higher $B$. The obtained $J^{\left[ 1\right] }$ with $\eta
_{11}\simeq 0.7$ and $\eta _{12}\simeq 0.3$ reproduces the data. The
analysis was not made for the data in Fig. 3(d) as the spin splitting is not
resolved at $\nu =8$. Consistency with the above $J^{\left[ 1\right] }$
values at $\nu =2$ and 4 for QPC2 supports the non-uniform heat
redistribution under heat conservation.

Here we comment on the large deviation of the observed $J$ from
the estimated $J^{\left[ 1\right] }$ in Fig. 3(b) with QPC1 showing
imperfect quantized conductance. Our analysis assumes that the heat
generated at QPC1 is equally split between the left- and right-going
downstream channels. However, in the presence of nonlinearity in the QPC1,
this is no longer true. Generally, electrons at higher energies are likely
to have higher transmission coefficients through the QPC barrier. Therefore,
the heat current can be increased up to $2J^{\left[ 1\right] }$ [the dashed
line in the bottom panel of Fig. 3(b)] with the $J^{\left[ 1\right] }$
estimated from the above procedure.

\begin{figure}[tbp]
\begin{center}
\includegraphics[width = 3.3in]{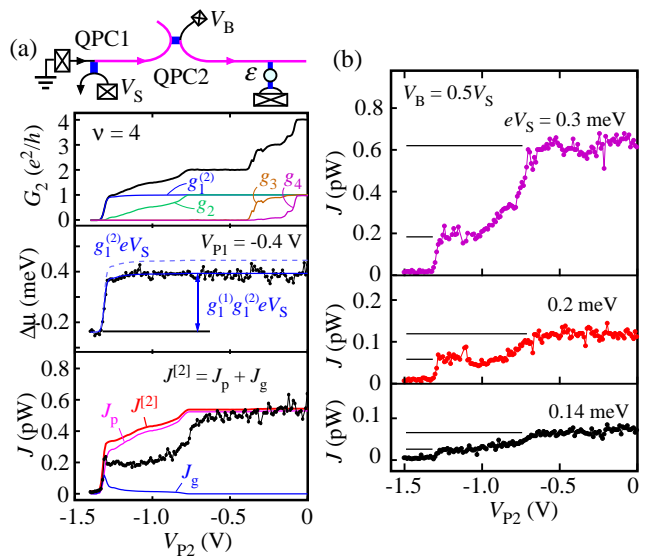}
\end{center}
\caption{Summary of double QPC experiments. (a) $G_{2}$ in the top panel
[same as in Fig. 3(a)], $\Delta \protect\mu $ in the middle panel, and the
heat current $J$ in the bottom panel. The data was obtained at $V_{\mathrm{B}%
}=0$ and $B=$ 1.8 T ($\protect\nu \simeq $ 4). The upper inset shows a
schematic channel configuration. (b) Heat current $J$ obtained at several $%
eV_{\mathrm{S}}$, where $V_{\mathrm{B}}=V_{\mathrm{S}}/2$ was chosen to
reduce the heat generation $J_{\mathrm{g}}$ at QPC2.}
\end{figure}

\subsection{Heat valve}

We demonstrate tunable heat transport through QPC2 as a heat valve. As shown
in Fig. 1(a) and the inset to Fig. 4(a), the two QPCs as well as the QD
spectrometer were activated. Heat is generated with QPC1 at $V_{\mathrm{P1}}$
= -0.4 V ($g_{1}^{(1)}\simeq $ 0.8, $g_{2}^{(1)}\simeq $ 0.3, and $%
g_{3}^{(1)}\simeq g_{4}^{(1)}\simeq 0$) and $\nu =4$, and should be
redistributed mostly in channels 1 and 2 before reaching QPC2. This heat is
partially transferred to the right segment through QPC2 and redistributed
again in the channels before reaching the QD detector. The heat current in
the right segment should be controlled by the conductance of QPC2 acting as
a heat valve.

Figure 4(a) summarizes $\Delta \mu $ in the middle panel and $J$ in the
bottom panel extracted from the QD spectroscopy, as well as $G_{2}$ in the
top panel. The measured $\Delta \mu $ (dots) follows $%
g_{1}^{(1)}g_{1}^{(2)}eV_{\mathrm{S}}$ (the solid line) with $g_{1}^{(1)}=$
0.8 and $g_{1}^{(2)}$ shown in the top panel. This indicates that the two
QPCs are working well in agreement with the characteristics obtained
separately. The heat current $J$ shows a double-step structure associated
with the heat partition between channels 1 and 2. This demonstrates the
heat-valve action with QPC2. In this experiment, the step heights in $J$
should depend on the heat fractionalization ratio $\eta _{ij}$. The first
and second steps associated with channels 1 and 2 have similar step heights,
as $\eta _{11}\simeq \eta _{12}$ in our device. The device should also
control the heat transport in channels 3 and 4, whereas corresponding step
structures are not visible due to the small $\eta _{13}$ and $\eta _{14}$ of
our device.

The expected heat current $J^{\left[ 2\right] }$ based on Eq. (\ref{JiR})
and $\eta _{ij}$ is plotted as a thick line together with the heat partition 
$J_{\mathrm{p}}$ [the first and the second terms of Eq. (\ref{JiR})] and
heat generation $J_{\mathrm{g}}$ (the third term) in the bottom panel of
Fig. 4(a). This suggests that the heat partition should be responsible for
the overall double-step structure in $J$. However, quantitative difference
between the calculated $J^{\left[ 2\right] }$ and measured $J$ is seen. The
observed stepwise feature of $J$ is much sharper than that of $J^{\left[ 2%
\right] }$ calculated from $G_{2}$, with the much wider first plateau. While
we do not have reasonable explanation for this, the difference might involve
the ignored many-body effect of the fractionalized quasiparticles.

The calculation suggests that the heat generation is significant only in the
first step where a finite bias $\mu _{1}=g_{1}^{(1)}eV_{\mathrm{S}}$ $\simeq 
$ 0.24 meV is applied across QPC2 in channel 1 for $V_{\mathrm{B}}$ = 0. One
can reduce the heat generation in channel 1 by adjusting $eV_{\mathrm{B}}$
at $\mu _{1}$, while this induces heat generation in other channels (finite $%
eV_{\mathrm{B}}$ against $\mu _{i}=0$ for $i\geq 2$). Considering that the
heat current is proportional to the square of the bias, a compromising
choice is choosing $V_{\mathrm{B}}=V_{\mathrm{S}}/2$. We also confirmed the
step-like control of $J$ for several $V_{\mathrm{S}}$ values with $V_{%
\mathrm{B}}=V_{\mathrm{S}}/2$, as shown in Fig. 4(b).

Unfortunately, the nonlinearity of the QPC conductance hinders further
quantitative analysis. For example, the QPC conductances $G_{1}$ and $G_{2}$
change slightly with the applied bias (not shown). The heat current even
without forming QPC2 at $V_{\mathrm{P2}}$ = 0 in Fig. 4(b) increases with $%
V_{\mathrm{S}}$ more rapidly as compared to the expected $V_{\mathrm{S}}^{2}$
dependence. Such non-linearity may change $\alpha $ and $\eta _{ij}$, and
cause some thermoelectric effects. This could be the reason why the ratio of
the first and second step heights in $J$ changes with $V_{\mathrm{S}}$. Such
non-linear effects as well as the many-body effects should be considered to
understand the observed behaviors.

\section{Summary}

In summary, heat transport through multiple QH edge channels is investigated
with a QPC heat injector, a QPC heat valve, and a QD thermometer in the QH
regime at $\nu $ = 2, 4, and 8 in an AlGaAs/GaAs heterostructure. The heat
fractionalization in the interacting channels is consistent with plasmon
transport model. The heat valve action is qualitatively consistent with the
single-particle tunneling at the QPC, while quantitative difference may
involve many-body effects and non-linear effects. The results encourage us
to study integrated heat circuits with QH edge channels.

\begin{acknowledgments}
We thank Shunya Akiyama for experimental supports. This work was supported
by JSPS KAKENHI JP19H05603 and Nanotechnology Platform Program at Tokyo
Institute of Technology.
\end{acknowledgments}

\appendix

\section{Heat redistribution}

The electronic heat is particle-hole excitations around the Fermi energy and
propagates in a form of charge density waves (plasmons) in the low energy
limit. We employ the distributed capacitance model to describe charge
density waves in multiple edge channels, as shown in Fig. 1(b). The charge
density $\boldsymbol{\rho }=\left\{ \rho _{i}\left( x,t\right) \right\} $,
the potential $\mathbf{V}=\left\{ V_{i}\left( x,t\right) \right\} $, and
time-dependent current $\mathbf{I}=\left\{ I_{i}\left( x,t\right) \right\} $
in vector representations for channel $i$ are defined as a function of
coordinate $x$ along the channels and time $t$. The electrostatic effects on
the occupation of the channels can be written in a simple form of $%
\boldsymbol{\rho }=\mathbf{CV}$ with the capacitance matrix $\mathbf{C}%
=\left\{ C_{ij}\right\} $ including self capacitance $C_{i}$ and
inter-channel capacitance $C_{ij}$. The quantized conductance $\sigma _{q}=%
\frac{e^{2}}{h}$ relates $\mathbf{I}=\sigma _{q}\mathbf{V}$. By using the
current conservation rule ($\frac{\partial }{\partial t}\boldsymbol{\rho }=-%
\frac{\partial }{\partial x}\mathbf{I}$), one can extract the unidirectional
wave equation of the form $\frac{\partial }{\partial t}\mathbf{I}=-\sigma
_{q}\mathbf{C}^{-1}\frac{\partial }{\partial x}\mathbf{I}$. Arbitrary waves
can be written with orthonormal eigenmode $\mathbf{\tilde{I}}_{m}=\left\{ 
\tilde{I}_{mi}\right\} $ and the velocity $v_{m}$ for mode index $m$, where $%
\mathbf{\tilde{I}}_{m}\cdot \mathbf{\tilde{I}}_{n}=\delta _{mn}$. Detailed
derivation and relation to the field theory are described in Refs. \cite%
{WashioPRB2016,FujisawaAnnPhys2022}. This scheme was successful in studying
charge dynamics in integer and fractional QH channels.\cite%
{Kamata-NatNano2014,Hashisaka-NatPhys2017,CJLinNatComm2021}

When a current fluctuation $I_{n}\left( x=0,t\right) $ is introduced, from a
QPC in the present case, at $x=0$ only in channel $n$, current fluctuation
at $x>0$ in channel $m$ is given by%
\begin{equation*}
I_{m}\left( x,t\right) =\sum\limits_{k}\tilde{I}_{km}\tilde{I}%
_{kn}I_{n}\left( 0,t-\frac{x}{v_{k}}\right) .
\end{equation*}%
The heat current $J_{m}\left( x\right) =\frac{1}{2}\sigma
_{q}^{-1}\left\langle \left( \Delta I_{m}\left( x,t\right) \right)
^{2}\right\rangle $ can be calculated from the fluctuating part $\Delta
I_{m}\left( x,t\right) =I_{m}\left( x,t\right) -\left\langle
I_{m}\right\rangle $ from the average $\left\langle I_{m}\right\rangle $,
and is given by%
\begin{equation*}
J_{m}\left( x\right) =\frac{1}{2}\sigma _{q}^{-1}\sum\limits_{k,l}\tilde{I}%
_{km}\tilde{I}_{kn}\tilde{I}_{lm}\tilde{I}_{ln}\phi \left( \frac{x}{v_{l}}-%
\frac{x}{v_{k}}\right) ,
\end{equation*}%
where $\phi \left( \Delta t\right) =\left\langle \Delta I_{n}\left(
0,t\right) \Delta I_{n}\left( 0,t-\Delta t\right) \right\rangle $ is the
correlation function of the input current fluctuation. We assume Lorentzian
spectrum with $\phi \left( \Delta t\right) =a\exp \left( -t/\tau \right) $,
where $a=\left\langle \left( \Delta I_{n}\left( 0,t\right) \right)
^{2}\right\rangle $ is the mean square of the input fluctuation and $\tau =%
\frac{\hbar }{eV_{\mathrm{S}}}$ is the correlation time. Then, the heat
current reaches the steady state at $x\gg $ $L_{\mathrm{F}}$, where the
fractionalization length $L_{\mathrm{F}}$ is defined as $L_{\mathrm{F}%
}=\max\limits_{m\neq n}\left\{ \frac{v_{m}v_{n}}{\left\vert
v_{m}-v_{n}\right\vert }\Delta t\right\} $. Therefore, the steady heat
current $J_{m}\left( \infty \right) $ at $x\rightarrow \infty $ can be
written as 
\begin{eqnarray*}
J_{m}\left( \infty \right) &=&\frac{1}{2}a\sigma
_{q}^{-1}\sum\limits_{k}\left( \tilde{I}_{km}\tilde{I}_{kn}\right) ^{2} \\
&=&\eta _{mn}J_{n}\left( 0\right) ,
\end{eqnarray*}%
where $J_{n}\left( 0\right) =\frac{1}{2}a\sigma _{q}^{-1}$ is the initial
heat current, and $\eta _{mn}=\sum\limits_{k}\left( \tilde{I}_{km}\tilde{I}%
_{kn}\right) ^{2}$ is the heat fractionalization factor discussed in this
paper. This form suggests reciprocal relation $\eta _{mn}=\eta
_{nm}$ in the heat redistribution.

For example, a system with two copropagating integer channels at $\nu =2$
can be written with eigenmodes%
\begin{equation*}
\mathbf{\tilde{I}}_{1}=\left( 
\begin{array}{c}
\cos \theta \\ 
\sin \theta%
\end{array}%
\right) \text{ and }\mathbf{\tilde{I}}_{2}=\left( 
\begin{array}{c}
\sin \theta \\ 
-\cos \theta%
\end{array}%
\right)
\end{equation*}%
by using mixing angle $\theta $ ($0\leq \theta \leq \pi /4$). Corresponding
heat fractionalization ratio reads 
\begin{eqnarray*}
\eta _{11} &=&\eta _{22}=\cos ^{4}\theta +\sin ^{4}\theta \\
\eta _{21} &=&\eta _{12}=1-\eta _{11}.
\end{eqnarray*}%
This allows us to calculate $\eta _{ij}$ from the charge fractionalization
data in previous reports ($\eta _{11}=$ 0.55 for $\theta =0.2\pi $ at $\nu =$
2).\cite{Hashisaka-NatPhys2017} Since the eigenmode $\mathbf{%
\tilde{I}}_{m}$ is determined by the capacitance matrix $\mathbf{C}$ of the
device, $\eta _{mn}$ should depend on the sample and may be controlled by
tailoring the electrostatic geometry.

\end{document}